\documentclass[12pt,preprint]{aastex}
\usepackage{spr-astr-addons}
\usepackage{lipsum,amsmath,multicol}
\usepackage{amsmath,amssymb}
\usepackage{graphicx,psfrag,subfigure}

\RequirePackage{color}

\begin{document}

\title{Charged anisotropic models for quark stars} 
\slugcomment{}
%% Running heads
\shorttitle{Short article title}
\shortauthors{Autors et al.}

%\email{\emaila}

\author{Jefta M. Sunzu} 
\affil{Astrophysics and Cosmology Research Unit, School of Mathematics, Statistics and Computer Science, University of KwaZulu-Natal, Private Bag X54001, Durban 4000, South Africa.\\
Permanent address: School of Mathematical Sciences, University of Dodoma, Tanzania.}

\author{Sunil D. Maharaj}
\affil{Astrophysics and Cosmology Research Unit, School of Mathematics, Statistics and Computer Science, University of KwaZulu-Natal, Private Bag X54001, Durban 4000, South Africa.}

\author{Subharthi Ray}
\affil{Astrophysics and Cosmology Research Unit, School of Mathematics, Statistics and Computer Science, University of KwaZulu-Natal, Private Bag X54001, Durban 4000, South Africa.}
%\maketitle

\begin{abstract} 
We perform a detailed physical analysis for a class of exact solutions for the Einstein-Maxwell equations. 
The linear equation of state consistent with quark stars has been incorporated in the model. The physical analysis of the exact solutions is 
performed by considering the charged anisotropic stars for the particular nonsingular exact model obtained by Maharaj, Sunzu and Ray. 
In performing such an analysis we regain masses obtained by previous researchers for isotropic and anisotropic matter. It is also indicated that 
other masses and radii may be generated which are in acceptable ranges consistent with observed values of stellar objects. A study of the mass-radius 
relation indicates the effect of the electromagnetic field and anisotropy on the mass of the relativistic star.
\end{abstract}

\section{Introduction}
The Einstein-Maxwell equations describe charged gravitating matter which are important in relativistic astrophysics,  and they model compact objects such 
as neutron stars, gravastars, dark energy stars and quark stars. In the study of such astrophysical compact objects, the Einstein-Maxwell field equations 
in static spherical spacetimes provide the basis of investigation, and have therefore attracted the attention of many researchers. With the help of these equations, 
researchers have discovered different structures and properties of relativistic stellar bodies relevant in astrophysical studies. For example, the 
solutions to the field equations in static spacetimes obtained by \cite{Thirukkanesh} describe realistic compact anisotropic 
spheres whose properties are relevant to stellar bodies such as SAXJ1808.4-3658. These solutions contain masses and central densities 
that correspond to realistic stellar bodies. Recently the models for charged matter generated by \cite{Farook} describe ultra compact astrophysical objects. 
The solutions obtained by \cite{Mehedi} describe charged compact objects and are compatible with well known stars. This indicates that 
the Einstein-Maxwell field equations have many applications in the study of relativistic astrophysical objects.

Pressure anisotropy is an important ingredient in many stellar systems in the absence of charge. Since the pioneering paper by \cite{Richard}, 
who were the first to consider pressure anisotropy in the study of anisotropic 
spheres in general relativity, there has been extensive research in this direction. It was established by \cite{Dev2} 
that pressure anisotropy has a significant effect on the structure and properties of stellar spheres. In particular it was shown that both the 
maximum mass and the redshift vary with the magnitude of the pressure anisotropy. For a positive measure of anisotropy, the stability of the 
sphere is enhanced when compared to isotropic configurations, and anisotropic distributions are stable for smaller adiabatic index 
values as shown by \cite{Dev} and \cite{Dev3}. Other uncharged anisotropic models with spherical 
symmetry include \cite{Mak3,Mak}, \cite{Harko}, \cite{Karmakar}, \cite{Mah-Chaisi,Maharaj4}, and 
\cite{Chaisi2,Chaisi,Chaisi3}. It is interesting to note the paper of \cite{Ivanov2} which showed that anisotropic 
models with heat flow can absorb the addition of charge, viscosity and convert null fluids to a perfect fluid.

It is important for many applications to include the electric field in stellar models. In particular, models with electric field permit 
causal signals over a wide range of parameters as illustrated by \cite{Sharma3}. It has been shown by \cite{Ivanov} 
that the presence of the electric field significantly affects the redshift, luminosity and mass of the compact object. Most of the models that include 
an electromagnetic field distribution are isotropic; these include the new classes of solutions obtained by \cite{Maharaj3}, 
\cite{Komathiraj2,Komathiraj3}, Thirukkanesh and Maharaj (2006, 2009) and 
\cite{Maharaj5}. Other stellar models that describe charged bodies with isotropic pressures are given by 
Chattopadhyay et al. (2012), Gupta and Maurya (2011a,b) and \cite{Naveen}. 
There are fewer research papers that include both anisotropic pressures and electromagnetic field distributions. The presence of pressure anisotropy 
with an electric field enhances the stability of a configuration under radial adiabatic perturbations compared to the matter with isotropic pressures. 
Stellar models containing both pressure anisotropy and electric field include compact objects admitting a one-parameter group of conformal motions 
of \cite{Esculpi}, the generalized isothermal models of \cite{Maharaj2}, the stellar models of 
\cite{Thirukkanesh}, and the regular compact models of \cite{Mafa}. Other charged anisotropic models are 
those of \cite{Farook} and \cite{Maurya}. However most of these models have the anisotropy parameter always 
present, and they do not contain isotropic solutions as a special case. It is important to build physical stellar models in which the 
anisotropy vanishes for an equilibrium configuration.
  
Different forms of the barotropic equation of state have been applied together with the field equations to find exact models that govern 
relativistic compactgravitating objects such as dark energy and quark strange stars (hybrid stars). \cite{Thirukkanesh3} have found exact solutions 
for the uncharged anisotropic sphere with the polytropic equation of state for particular 
choices of the polytropic index. \cite{Mafa2} used the same general polytropic equation of state, and obtained exact 
solutions for field equations in the presence of the electromagnetic field and anisotropic pressures. \cite{Shibata} studied the stability of 
rotating bodies and \cite{Lai} indicated that large amounts of gravitational energy are released in the gravitational collapse of polytropes. 
Other treatments on polytropes include the results of \cite{Tooper}, \cite{Nilsson}, \cite{Mach} and \cite{Heinzle}. \cite{Maharaj} and Feroze and Siddiqui (2011)
found exact solutions of the Einstein-Maxwell field equations for 
charged anisotropic stars using a quadratic equation of state. There have been many anisotropic and charged exact models with a linear equation of state: 
\cite{Mafa} generated compact exact models with regular distributions, \cite{Thirukkanesh} found models 
consistent with dark energy stars and quark stars, \cite{Maharaj2} generated anisotropic isothermal models, 
\cite{Sharma} found models consistent with quark matter, and \cite{Esculpi} generated conformally invariant spheres. 
However, in general, most of these models do not regain charged isotropic models. Some analytical solutions to the field equations with a 
linear quark equation of state for charged isotropic stars were found by \cite{Komathiraj}. Using the same equation of state, 
\cite{Sotani1}, \cite{Sotani2}, and \cite{Bombaci} analysed quark stars with isotropic pressures. 
There has been an extension of the linear quark equation to include anisotropic pressures in modeling the behaviour of strange stars by 
\cite{Farook}, \cite{Mehedi1}, and \cite{Mak5}.

It should be noted that the microscopic effects of the strange quark matter coming from strong 
interactions of QCD (e.g., see \cite{dong13} and  \cite{Dey}) are all encrypted in the final form in the equation 
of state of matter. We study the general relativistic behaviour of these equations of states, by employing a linear 
approximation for these strange matter equation of states. Such approximations of the equation of states can
 be found in the literature in the study of various properties of compact stars. The linear approximation 
 of the strange quark matter equation of state has been used by \cite{zdunik2000} to study the quasi periodic
  oscillation (QPO) frequencies in the Lower Mass X-ray Binaries (LMXBs). \cite{dorota2000} used the linear approximation of strange matter to compute the mass shedding limit of strange stars. 

Recently a class of exact isotropic solutions of Einstein's equations for non-rotating relativistic 
stars has also been studied by \cite{murad2014}. They also comment that as strange 
stars are not purely gravitationally bound; they are bound by strong interactions. 
Study of the same in the light of modified gravity theories should not produce any 
difference in the mass-radius relation. In this context, although \cite{salvatore2013} showed that there is an increase in the mass of neutron stars in the $f(R)=R+R\left(e^{-R/R_0} - 1\right) $ gravity model, \cite{ganguly2014}  showed that for the $f(R)=R+\alpha R^2$ model (and subsequently many other $f(R)$ models where the uniqueness theorem is valid) the existence of
compact compact astrophysical is highly unnatural. This is because
the equation of state of a compact star should be completely determined by
the physics of nuclear matter at high density, and not only by the theory of gravity.

The objective of this paper is to perform a detailed physical analysis of the particular exact solutions to the Einstein-Maxwell system of 
equations with a linear quark equation of state for charged anisotropic stars obtained by \cite{Jefta1}. In performing 
such a physical analysis we seek to regain models with masses and radii obtained by other researchers, and show that other masses contained in our model 
are in acceptable ranges. We also seek to compare masses and radii by considering anisotropic and isotropic pressures. This analysis shows that the relevant 
class of exact solutions with a quark equation of state has astrophysical significance.

\section{The model \label{one}}
We model the stellar interior with quark matter in general relativity. The spacetime geometry is spherically symmetric and given by
\begin{equation}
 ds^{2}=-e^{2\nu(r)}dt^{2}+e^{2\lambda(r)}dr^{2}+r^{2}(d\theta^{2}+\sin^{2}\theta d\phi^{2}),
\label{line-element}
\end{equation}
where $\nu(r)$ and $\lambda(r)$ are the gravitational potentials. The Reissner-Nordstrom line element
\begin{eqnarray}
&&ds^{2}= \nonumber \\
&&-\left(1-\frac{2M}{r}+\frac{Q^{2}}{r^{2}}\right)dt^{2}  +\left(1-\frac{2M}{r} +\frac{Q^{2}}{r^{2}}\right)^{-1}dr^{2} \nonumber\\
&&+ r^{2}(d\theta^{2}+\sin^{2}\theta d\phi^{2}),
\label{line-element-exterior}
\end{eqnarray}
describes the exterior spacetime. The quantities $M$ and $Q$ define the total mass and charge of the star, respectively.
The energy momentum tensor is defined by
\begin{eqnarray}
 T_{ij}&=&\mbox{diag}\left(-\rho-\frac{1}{2}E^{2},p_{r}-\frac{1}{2}E^{2},p_{t}+\frac{1}{2}E^{2},\right. \nonumber\\
 && \left. p_{t}+\frac{1}{2}E^{2}\right),
\label{Energy-mom tensor}
\end{eqnarray}
in the presence of charge and anisotropy. The energy density ($\rho$), the radial pressure ($p_{r}$), the tangential pressure ($p_{t}$), 
and the electric field intensity ($E$) are measured relative to a comoving fluid four-velocity $u^{a}(u^{a}u_{a}=-1)$.

The Einstein-Maxwell field equations are given by
\begin{subequations}
\label{Emf}
\begin{eqnarray}
 \dfrac{1}{r^{2}}\left( 1-e^{-2\lambda}\right)+\dfrac{2\lambda^\prime}{r}e^{-2\lambda}&=&\rho + \frac{1}{2}E^{2}, \\
\label{Emf1}
 -\dfrac{1}{r^{2}}\left( 1-e^{-2\lambda}\right)+\dfrac{2\nu^{\prime}}{r}e^{-2\lambda}&=&p_{r} - \frac{1}{2}E^{2},\\
\label{Emf2}
e^{-2\lambda}\left( \nu^{\prime\prime}+\nu^{\prime^{2}}-\nu^{\prime}\lambda^{\prime} \right. && \nonumber\\
\left. +\dfrac{\nu^{\prime}}{r}-\dfrac{\lambda^{\prime}}{r}\right) &=& p_{t}+ \frac{1}{2}E^{2},\\
\label{Emf3}
 \sigma&=&\frac{1}{r^{2}}e^{-\lambda}\left( r^{2}E\right)^{\prime},
\label{Emf4}
\end{eqnarray}
\end{subequations}
where primes denote differentiation with respect to the radial coordinate $r$. The function $\sigma$ represents the proper charge density. The equation of state is linear and of the form
\begin{equation}
p_{r}=\frac{1}{3}\left(\rho-4B\right),
\label{eqnstate}
\end{equation}
where $B$ is a constant related to the surface density of the stellar body representing a sharp surface. If we consider the MIT bag model for quark stars, 
then $B$ can also be identified with the bag constant.

We introduce a new independent variable $x$ and define the metric functions
$Z(x)$ and $y(x)$ as
\begin{equation}
 x=Cr^{2},\;\;Z(x)=e^{-2\lambda(r)},\;\;A^{2}y^{2}(x)=e^{2\nu(r)},
\label{transformation}
\end{equation}
where $A$ and $C$ are arbitrary constants following \cite{Durgapal}. 
With this transformation the line element in (\ref{line-element}) becomes
\begin{equation}
 ds^{2}=-A^{2}y^{2}dt^{2}+\frac{1}{4xCZ}dx^{2}+\frac{x}{C}(d\theta^{2}+\sin^{2}\theta d\phi^{2}).
\label{newlineelement}
\end{equation}
The Einstein-Maxwell field equations become 
\begin{subequations}
\label{nnewemf}
\begin{eqnarray}
 \rho&=&3p_{r}+4B, \label{nnewemf1}\\
 \frac{p_{r}}{C}&=&Z\frac{\dot{y}}{y}-\frac{\dot{Z}}{2}-\frac{B}{C}, \label{nnewemf2}\\
  p_{t}&=&p_{r}+\Delta,\label{nnewtangetial}\\
 \Delta &=&\frac{4xCZ\ddot{y}}{y}+C\left(2x\dot{Z}+6Z\right)\frac{\dot{y}}{y}\nonumber \\
 & & +C\left(2\left(\dot{Z}+\frac{B}{C}\right)+\frac{Z-1}{x}\right),\label{nnewemf4}\\ 
\frac{E^{2}}{2C}&=&\frac{1-Z}{x}-3Z\frac{\dot{y}}{y}-\frac{\dot{Z}}{2}-\frac{B}{C},\label{nnewemf3}\\ 
\sigma&=& 2\sqrt{\frac{ZC}{x}}\left(x\dot{E}+E\right),\label{nnewemf5}
\end{eqnarray}
\end{subequations}
where dots represent derivatives with respect to the variable $x$. The quantity $\Delta= p_{t}-p_{r}$ is called the measure of anisotropy. 
We introduce the mass function given by
\begin{equation}
 M(x)=\frac{1}{4C^\frac{3}{2}}\int_{0}^{x}\sqrt{\omega}\left(  \rho_{*}+E^{2} \right) d\omega,
\label{mass2}
\end{equation}
where
\begin{equation}
\rho_{*}= \left( \frac{1-Z}{x}-2\dot{Z}\right )C,
\end{equation}
is the energy density when the electric field $E=0$.

Some solutions to the system (\ref{nnewemf}), applicable to quark matter, were presented in \cite{Jefta1}. In that model it was assumed that: 
\begin{eqnarray*}
y&=&\left(a+x^{m}\right)^{n}, \label{Choice-y}\\
\Delta &=&A_{0}+A_{1}x+A_{2}x^{2}+A_{3}x^{3}.
\label{choice-delta}
\end{eqnarray*}
For particular choices of the parameters $m$ and $n$ it is possible to integrate the Einstein-Maxwell system exactly. 
The choice of anisotropy ensures isotropic pressures can be regained. To ensure that the anisotropy vanishes at the stellar centre 
we should set $A_{0}=0$. Here we consider a particular solution of \cite{Jefta1} that enables us to perform a detailed physical 
analysis. The particular solution that we can utilize can be written in terms of analytical functions, and this is presented in the 
Appendix. Note that this generalized class of models with a quark equation of state contains the nonsingular solutions of 
\cite{Komathiraj} with isotropic pressures.

\begin{table*}
\caption{Various parameter values for various models of stellar objects}
\begin{tabular}{|c|c|c|c|c|c|c|c|c|c|}
 \hline
$\tilde{a}$ &$\tilde{B}$&$\tilde{C}$&$\tilde{A_{1}}$&$\tilde{A_{2}}$&$\tilde{A_{3}}$& radius
& mass & Model\\
\hline
$52$ & $12$ & $1$ & $20$ & $25$ & $20$ & $9.46$ & $2.86M_{\odot}$ & \cite{Mak4}\\
\hline
$350$ & $12$ & $1$  & $250$ & $280$ & $290$ & $10.99$ & $  2.02M_{\odot}$ & \cite{Rodrigo} \\
\hline
$350$ & $12$ & $1$  & $230$ & $235$ & $240$ & $9.40$ & $ 1.67M_{\odot}$ &  \cite{Taparati}\\
\hline
$202$ & $12$ & $1$  & $25$ & $20$ & $20$ & $7.07$ & $ 1.433M_{\odot}$ &  \cite{Dey}\\
\hline
$350$ & $12$ & $1$ & $289$ & $200$ & $260$ & $7.07$ & $0.94M_{\odot}$ & \cite{Thirukkanesh}\\
\hline
\end{tabular}
\label{table0}
\end{table*}

\begin{table*}
\caption{Masses and radii for isotropic and anisotropic stars for different choice of parameters}
\begin{tabular}{|c|c|c|c|c|c|c|c|c|c|c|}
 \hline
Name &$\tilde{a}$ &$\tilde{B}$&$\tilde{C}$&$\tilde{A_{1}}$&$\tilde{A_{2}}$&$\tilde{A_{3}}$&$r_{(\Delta \neq 0)}$&$r_{(\Delta = 0)}$ 
& $\left(\frac{M}{M_{\odot}}\right)_{\Delta \neq 0}$ & $\left(\frac{M}{M_{\odot}}\right)_{\Delta = 0}$\\
\hline
R1&$285$ & $12$ & $1$  & $25$ & $20$ & $25$ & $6.84$ & $6.85$ & $1.28994$ & $1.31530$\\
\hline
R2&$100$ & $12$ & $1$  & $20$ & $5$ & $10$ & $6.67$ & $6.68$ & $1.56259$ & $1.56730$\\
\hline
R3&$260$ & $10$ & $1$ & $35$ & $25$ & $30$ & $7.59$ & $7.61$ & $1.58585$ & $1.61878$\\
\hline
R4&$260$ & $10$ & $1$ & $20$ & $30$ & $20$ & $7.60$ & $7.61$ & $1.60033$ & $1.61878$\\
\hline
R5&$200$ & $10$ & $1$  & $40$ & $30$ & $40$ & $7.57$ & $7.59$ & $1.66749$ & $1.69064$\\
\hline
R6&$35$ & $12$ & $1$ & $25$ & $10$ & $15$ & $5.78$ & $5.77$ & $1.73268$ & $1.72885$\\
\hline
\end{tabular}
\label{tableone}
\end{table*}

\begin{table*}
\caption{Variation of parameter $\tilde{A_{1}}$ for $\tilde{a}=260$, $\tilde{B}=10$, $\tilde{C}=1$, $\tilde{A_{2}}=15$, $\tilde{A_{3}}=20$}
\begin{tabular}{|c|c|c|c|c|}
 \hline
$\tilde{A_{1}}$& $r_{(\Delta \neq 0)}$ & $r_{(\Delta = 0)}$ & $\left(\frac{M}{M_{\odot}}\right)_{\Delta \neq 0}$ 
& $\left(\frac{M}{M_{\odot}}\right)_{\Delta = 0}$\\
\hline
$5$  & $7.6100$ &  & $1.61456$ & \\
%\hline
$10$  & $7.6100$ &  & $1.61087$ & \\
%\hline
$15$ & $7.6100$ &  & $1.60718$ & \\
%\hline
$20$ &  $7.6100$ & $7.6100$ & $1.60349$ & $1.61878$\\
%\hline
$25$ & $7.6100$ & & $1.59981$ & \\
%\hline
$30$ & $7.6100$ & & $1.59612$ & \\
%\hline
$35$ & $7.6100$ & & $1.59243$ & \\
%\hline
$40$ & $7.6100$ & & $1.58874$ & \\
\hline
\end{tabular}
\label{table2}
\end{table*}

\section{Stellar masses}
Our exact solutions are more general than earlier treatments and have the flexibility of allowing for the fine-tuning of the parameters. 
The right choice of parameters in the multi-dimensional parameter space enables us to  regain the stellar masses of compact bodies 
previously identified by many other research groups. To start with, we make the following transformations:
 \begin{eqnarray} && \tilde A_{1}= A_{1}R^{2}, \, \tilde A_{2}=A_{2}R^{2}, \, \tilde A_{3}=A_{3}R^{2},  \nonumber \\
&&  \tilde B= BR^{2}, \, \tilde C=CR^{2}, \,  \tilde a=aR^{2}, \nonumber
\end{eqnarray}
where $R$ takes the same unit as $x$, and in order to match with the realistic units, it is renormalised by a factor of 43.245, i.e., $R=43.245 x$.
In the literature, we find many observed and analysed charge-neutral compact star masses, varying from 0.9~$M_\odot$ to 2.01~$M_\odot$. 
The studies of charged stars however allow for more mass in the stable configuration. In our present study, we aim to regain masses of some of the observed 
compact stellar bodies for the uncharged cases identified to be strange stars, thereby narrowing the  parameter ranges in our model for the 
existence of such objects. For the charged cases, we follow 
the same exercise to regain the values of the theoretically obtained masses for the charged stars.

In particular,  for the electrically charged strange quark stars, we have  regained the mass $M=2.86M_{\odot}$ with radius $r=9.46$ km consistent 
with mass and radius obtained by \cite{Mak4}, the mass $M= 2.02 M_{\odot}$ with radius $r=10.99$ km consistent with the object found by  
\cite{Rodrigo}, and the mass $M=1.433M_{\odot}$ with radius $r=7.07$ km
consistent with the particular results obtained by \cite{Thirukkanesh} and Mafa Takisa and Maharaj (2013a).

Charged compact stars have been identified as quark stars: the mass $M=1.67M_{\odot}$ with 
radius of $9.4$ km consistent with the star PSR J1903+327 is discussed by \cite{Freire} and 
\cite{Taparati}, and the mass $M=1.433M_{\odot}$ with radius of $7.07$ km was found by 
\cite{Dey} in their strange star models. Parameter values which 
give these masses and radii in our model are given in Table \ref{table0}.

\section{Physical analysis}
Although numerically  we could regain the values of masses and radii of many previously obtained stellar models, a systematic study of the variation of the 
anisotropic parameters in our model is also necessary. To this end, we study the effect of anisotropic parameters $\tilde A_{1}$, $\tilde A_{2}$ and $\tilde A_{3}$ 
on masses and radii of stellar bodies, by varying one parameter at a time and keeping the others fixed. Also, to make the effects more pronounced, 
we have chosen a few sets of parameters so as to give a value of the mass-radius relation in the \emph{acceptable} range. The surface of the anisotropic 
star is considered to be a point of vanishing radial pressure. Also, in our most general solution we have set $\tilde{A_{1}}=\tilde{A_{2}}=\tilde{A_{3}}=0$ 
so as to obtain the isotropic model.

Table \ref{tableone} shows different masses and radii for isotropic and anisotropic stars for different choices of parameters. This study shows that the 
masses and radii of the anisotropic stars are less than the corresponding quantities for isotropic stars for most of the values of the parameters chosen. 
However, the model indicates that there are also some values of the parameters which give the mass and radius of the anisotropic star greater 
than the corresponding value for the isotropic star. This is shown in the last row (R6) which indicates small values of radii but greater masses 
for both anisotropic and isotropic stars. The masses and radii indicated in Table \ref{tableone} are in the acceptable range for the quark stars as studied by 
\cite{Taparati}.

The effect of the parameter $\tilde{A_{1}}$ on the mass and radius of the anisotropic star is shown in Table \ref{table2}. 
As $\tilde {A_{1}}$ increases the mass of the anisotropic star decreases while the radius remains constant. The corresponding mass and radius for 
the isotropic case are  $1.61878 {M_{\odot}}$ and $7.61$ km respectively.

However, variation of the parameter $A_{2}$ for fixed values of $\tilde{a}=260$, $\tilde{B}=10$, $\tilde{C}=1$, $\tilde{A_{1}}=20$, $\tilde{A_{3}}=20$, did not result in any visible change of the mass and the radius. Carrying out the variation  of $A_2$ up to two orders of magnitude (from 5 to 100), we have found that the mass of the anisotropic star is constant 
around $1.60033M_{\odot}$ with radius $7.60$ km while the isotropic star has the mass $1.61878 M_{\odot}$ with radius of $7.61$ km. The difference in the mass from the isotropic to the anisotropic case here is due to the presence of the  other constant anisotropic parameters $A_{1}$ and  $A_{3}$.

\begin{table}
\caption{Variation of parameter $A_{3}$ for $\tilde{a}=260$, $\tilde{B}=10$, $\tilde{C}=1$, $\tilde{A_{1}}=20$, $\tilde{A_{2}}=15$}
\begin{tabular}{|c|c|c|c|c|}
 \hline
$\tilde{A_{3}}$& $r_{(\Delta \neq 0)}$ & $r_{(\Delta = 0)}$ & $\left(\frac{M}{M_{\odot}}\right)_{\Delta \neq 0}$ 
& $\left(\frac{M}{M_{\odot}}\right)_{\Delta = 0}$\\
\hline
$5$  & $7.6000$ & $ $ & $1.60073$ & $ $\\
%\hline
$10$  & $7.6000$ & $ $ & $1.60060$ & $ $\\
%\hline
$15$ & $7.6000$ & $ $ & $1.60046$ & $ $\\
%\hline
$20$ &  $7.6000$ & $7.6100$ & $1.60033$ & $1.61878$\\
%\hline
$25$ & $7.6000$ & $ $ & $1.60020$ & $ $\\
%\hline
$30$ & $7.6000$ & $ $ & $1.60006$ & $ $\\
%\hline
$35$ & $7.6000$ & $ $ & $1.59993$ & $ $\\
%\hline
$40$ & $7.6000$ & $ $ & $1.59980$ & $ $\\
%\hline
$100$ & $7.6000$ & $ $ & $1.59820$ & $ $\\
\hline
\end{tabular}
\label{table4}
\end{table}
In Table \ref{table4}, we see that the variation of the parameter $\tilde{A}_{3}$ does not affect the radius of the star. 
However, there is a decrease in the mass of the star  with the increase of $\tilde{A_{3}}$.  This decrease in the mass however still falls in the allowed range of the masses and radii for quark stars.

One noticeable feature of our results is that with the increase of the anisotropic parameters $\tilde{A}_{1}$ and $\tilde{A}_{3}$, there is decrease in the mass of the star, whereas the radius remain constant. This essentially means that the effective density of the star decreases with the increase of the anisotropic parameters, and hence the effective equation of state become stiffer.

\begin{figure}
\centering
 \includegraphics[width=6cm,height=8cm, angle=-90]{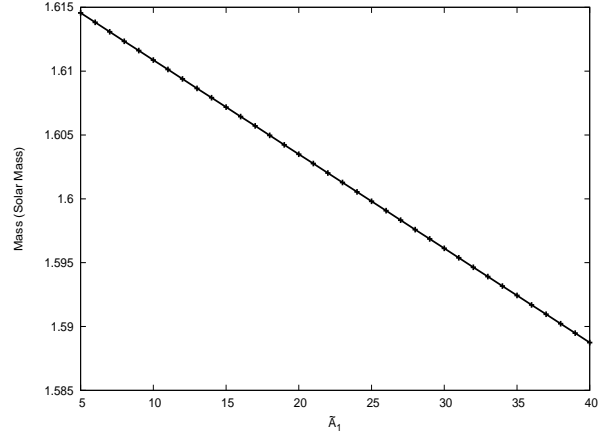}
 \caption{Variation of the mass with the parameter $\tilde{A}_{1}$, keeping other parameters constant}
\label{figA1}
\end{figure}

\begin{figure}
\centering
\includegraphics[width=6cm,height=8cm, angle=-90]{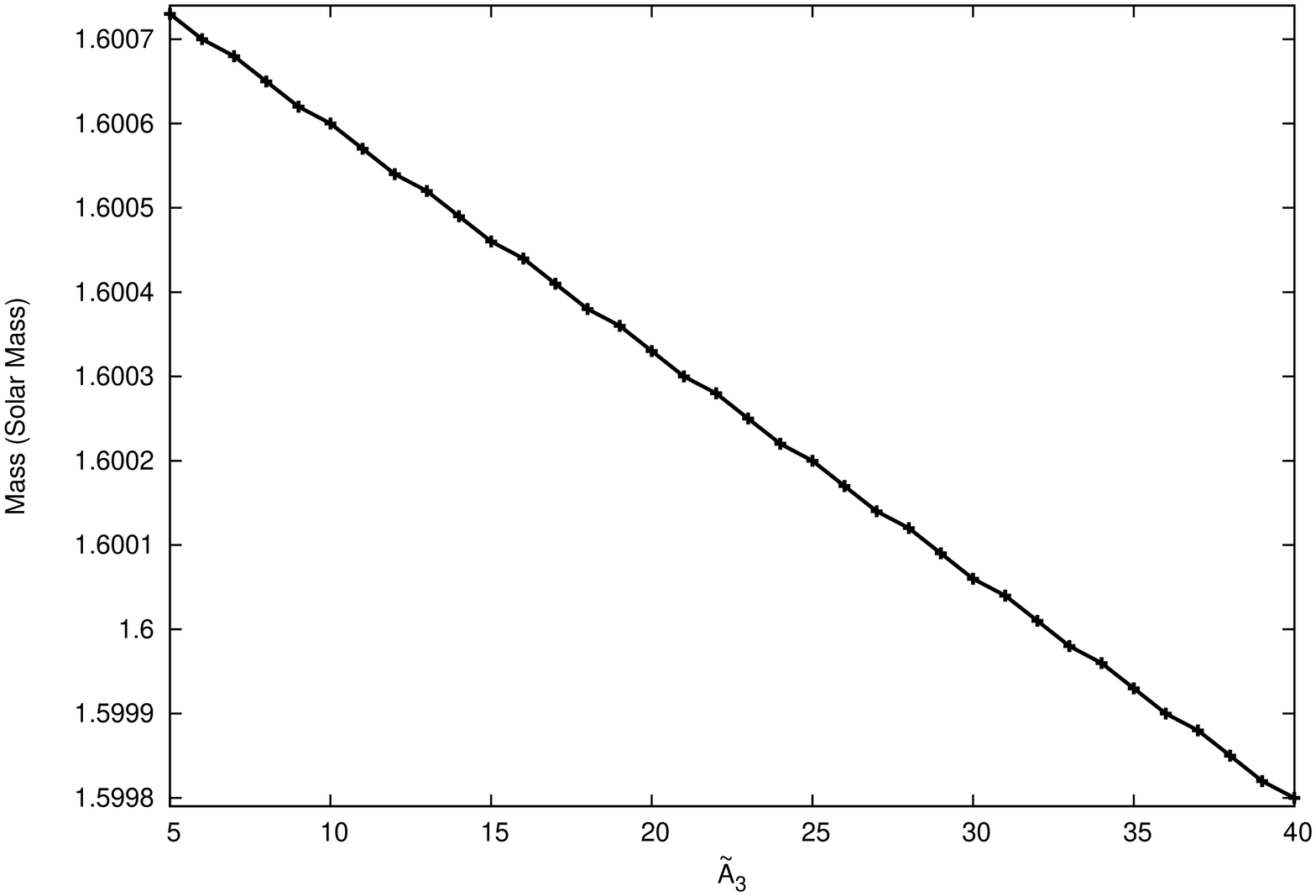}
\caption{Variation of the mass with the parameter $\tilde{A}_{3}$, keeping other parameters constant}
\label{figA3}
\end{figure}

\begin{figure}[h!]
\centering
\includegraphics[width=8cm]{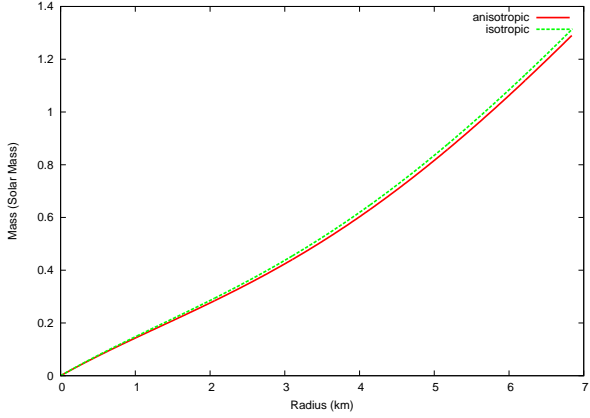}
\caption{ The mass-radius relation using parametric values indicated by $R1$. Here we see that there is an increase in the values for 
the isotropic case as compared to the anisotropic ones} 
 \label{nine}
\end{figure}

\begin{figure}[h!]
 \centering
 \includegraphics[width=8cm]{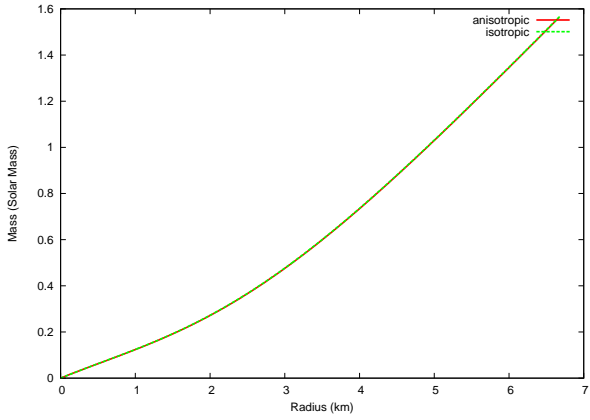}
 \caption{The mass-radius relation using parametric values indicated by $R2$. Clearly, for these choices of parameter sets, 
there is not much difference between the anisotropic and the isotropic cases} 
 \label{ten}
\end{figure}

\begin{figure}[h!]
 \centering
 \includegraphics[width=8cm]{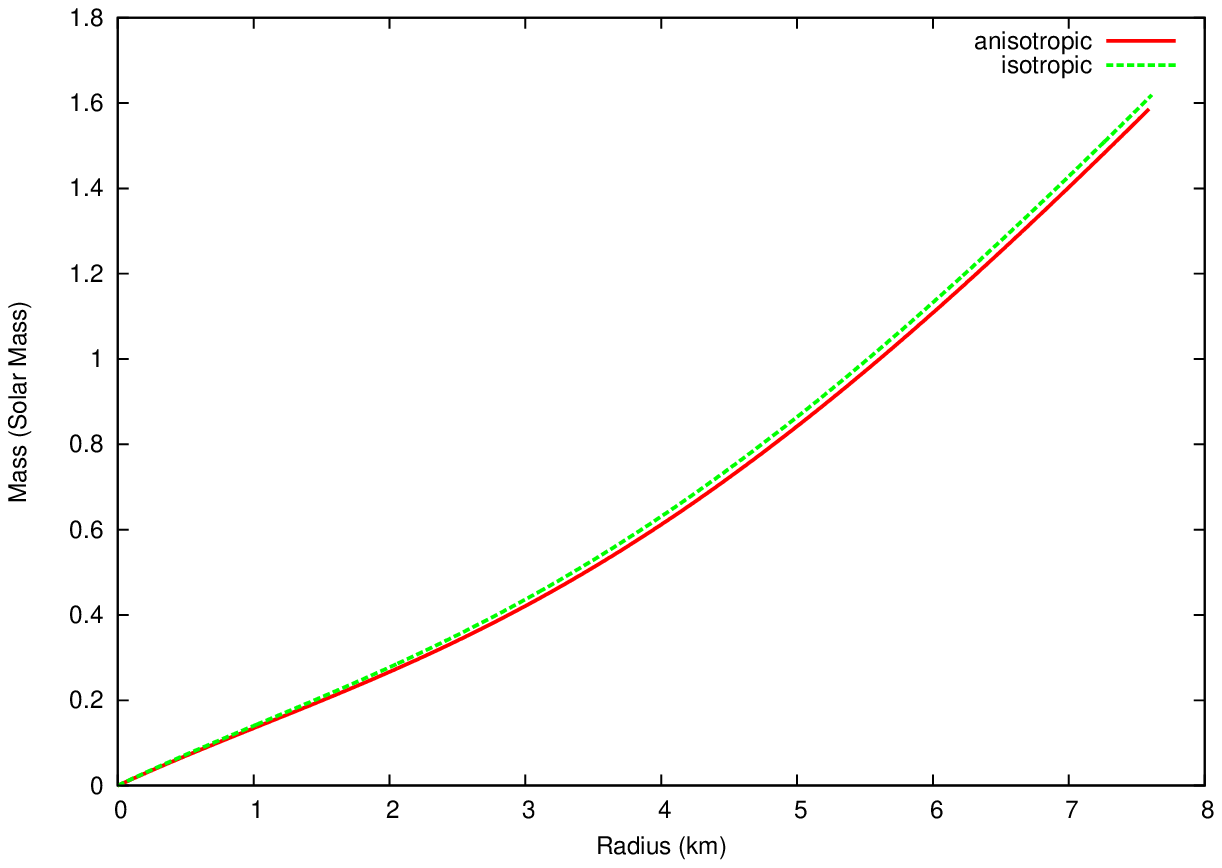}
 \caption{ The mass-radius relation using parametric values labelled by $R3$} 
 \label{eleven}
\end{figure}
\begin{figure}[h!]
 \centering
 \includegraphics[width=8cm]{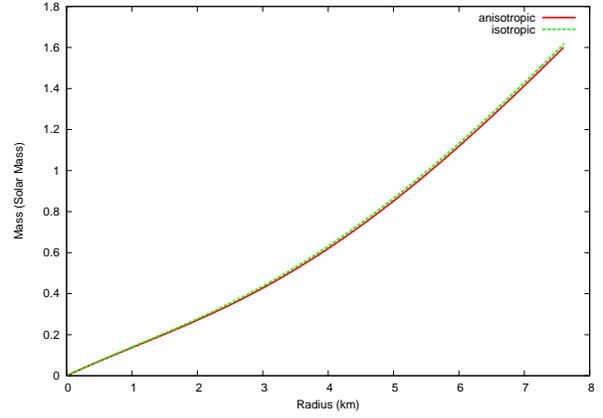}
 \caption{The mass-radius relation using parametric values indicated by $R4$} 
 \label{twelve}
\end{figure}

\begin{figure}[h!]
 \centering
 \includegraphics[width=8cm]{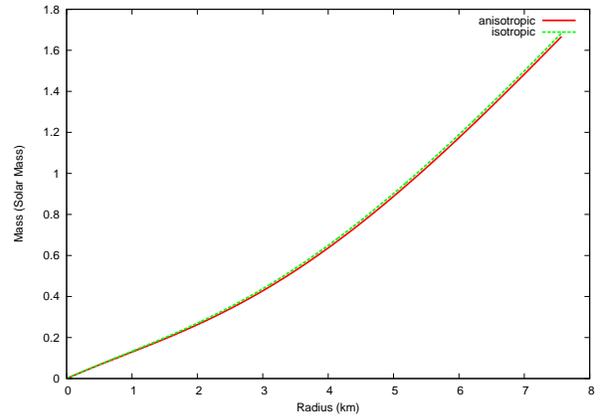}
 \caption{The mass-radius relation using parametric values given by $R5$}
 \label{thirteen}
\end{figure}

\begin{figure}[h!]
 \centering
 \includegraphics[width=8cm]{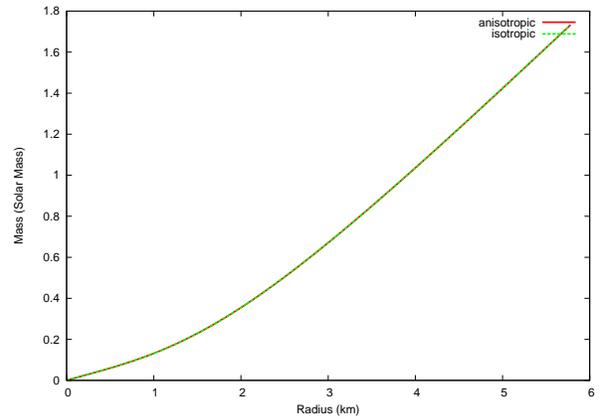}
 \caption{The mass-radius relation using parametric values labelled by $R6$} 
 \label{fourteen}
\end{figure}
In Fig. \ref{figA1} and Fig. \ref{figA3} we learn that the mass of the star decreases linearly with an increase of the anisotropic parameters $\tilde{A_{1}}$ 
and $\tilde{A_{3}}$ in the identified range of the constants. In order to compare the variation of the mass throughout the interior of anisotropic and isotropic stars, 
we have plotted in Fig. \ref{nine}-\ref{fourteen} the mass against radial distance using the parameter values in Table \ref{tableone}. 
In general, we see that the masses for the  isotropic cases are larger than their anisotropic counterparts, except for the parameter sets R2 and R6, 
where the two graphs appear to overlap.

\section{Conclusion}
We have studied the exact class of models with a quark equation of state found by \cite{Jefta1}. 
A detailed analysis of the physical features of this class was performed. In the uncharged case we regained the masses of \cite{Dey} and 
\cite{Taparati}. For the charged case we regained the models of \cite{Mak4}, \cite{Rodrigo}, and  
\cite{Thirukkanesh}. For the anisotropic case we have also generated masses ranging from $1.28994M_{\odot}$ to $1.73268M_{\odot}$ with radius of the range 
$5.78$ km to $7.61$ km. For the isotropic case the masses generated are from $1.31530 M_{\odot}$ to $1.72885 M_{\odot}$ with radius varying from $5.77$ km to 
$7.61$ km. These particular models are good candidates for the astrophysical object SAXJ1808.4-3658. As the form of the measure of anisotropy depends 
on the parameters $A_{1}$, $A_{2}$, $A_{3}$ it is possible to generate a wide variety of other compact objects for a suitable choice of parameters. A detailed study of the mass-radius 
relation for different anisotropy parameter values highlights the effect of the anisotropy on the mass of the star.

\acknowledgments
We are grateful to the National Research Foundation and the University of KwaZulu-Natal for financial support. 
JMS extends his appreciation to the University of Dodoma in Tanzania for study leave. SDM acknowledges that this work is 
based upon research supported by the South African Research Chair Initiative of the Department of Science and Technology and the National 
Research Foundation.

\appendix
\section*{Appendix}\label{appendix-model-solution}
In this Appendix we consider an exact solution to the Einstein-Maxwell system (\ref{nnewemf}) found by  \cite{Jefta1}. This class of 
exact solutions can be written in terms of elementary functions and is given by:
\begin{subequations}
\begin{eqnarray*}
 e^{2\nu}&=&A^{2}\left(a+x\right)^{4},\label{nonsingular-potential1}\\
 e^{2\lambda}&=&\dfrac{{315( a+x)^{2}(a+3x)}}{9\left( 35a^{3}+35a^{2}x+21ax^{2} + 5x^{3} \right)
-\dfrac{2Bx}{C}\left( 105a^{3}+189a^{2}x+135ax^{2} + 35x^{3} \right)
+\frac{315 L(x)}{C}},\label{nonsingular-potential2}\\
\rho&=&\dfrac{3C\left( 140a^{4}+434a^{3}x+318a^{2}x^{2} + 150ax^{3}+30x^{4}\right)}{35( a+x)^{3}(a+3x)^{2}}\nonumber\\
& &+\dfrac{3\Psi(x)+B\left( 210a^{5}+798a^{4}x+1476a^{3}x^{2} + 2540a^{2}x^{3}+2090 ax^{4} +630 x^{5} \right) }{105( a+x)^{3}(a+3x)^{2}},\label{nonsingular-energydensity}\\
p_{r}&=&\dfrac{C\left( 140a^{4}+434a^{3}x+318a^{2}x^{2} + 150ax^{3}+30x^{4}\right)}{35( a+x)^{3}(a+3x)^{2}}\nonumber\\
& &+\dfrac{\Psi(x)-B\left( 70a^{5}+994a^{4}x+3708a^{3}x^{2} + \frac{16780}{3}a^{2}x^{3}+\frac{11770}{3}ax^{4} +1050x^{5} \right) }{105( a+x)^{3}(a+3x)^{2}},\label{nonsingular-radialp}\\
 p_{t}&=&\dfrac{C\left( 140a^{4}+434a^{3}x+318a^{2}x^{2} + 150ax^{3}+30x^{4}\right)}{35( a+x)^{3}(a+3x)^{2}}\nonumber\\
& & +\dfrac{\Omega(x)-B\left( 70a^{5}+994a^{4}x+3708a^{3}x^{2} + \frac{16780}{3}a^{2}x^{3}+\frac{11770}{3}ax^{4} +1050x^{5} \right) }{105( a+x)^{3}(a+3x)^{2}},\label{nonsingular-tangentialp}\\
\Delta&=&A_{1}x+A_{2}x^{2}+A_{3}x^{3},\\
 E^{2}&=&\dfrac{C\left(1764a^{3}x+13068a^{2}x^{2}+12204ax^{3} + 3780x^{4}\right)-\Lambda(x)}{315( a+x)^{3}(a+3x)^{2}}\nonumber\\
& &-\dfrac{B\left(168a^{4}x+1296a^{3}x^{2}+6528a^{2}x^{3}+7280ax^{4}+2520x^{5}\right) }{315( a+x)^{3}(a+3x)^{2}},
\label{nonsingular-electricf}
\end{eqnarray*}
\label{nonsingular-exact}
\end{subequations}
where
\begin{eqnarray*}
 L(x)&=&A_{1}\left( \frac{1}{5}a^{3}x^{2}+\frac{3}{7}a^{2}x^{3}+\frac{1}{3}ax^{4}+\frac{1}{11}x^{5}\right)\\
 & &+A_{2}\left( \frac{1}{7}a^{3}x^{3}+\frac{1}{3}a^{2}x^{4}+\frac{3}{11}ax^{5}+\frac{1}{13}x^{6}\right)\\
 & &+A_{3}\left(  \frac{1}{9}a^{3}x^{4}+\frac{3}{11}a^{2}x^{5}+\frac{3}{13}ax^{6}+\frac{1}{15}x^{7}\right),\\
\Psi(x) &=&A_{1}x\left(-21a^{5}-57a^{4}x+20a^{3}x^{2}+\frac{1360}{11}a^{2}x^{3}+105ax^{4} +\frac{315}{11}x^{5} \right)\\
& & +A_{2}x^{2}\left(-\frac{45}{2}a^{5}-\frac{185}{2}a^{4}x-\frac{1145}{11}a^{3}x^{2}-\frac{315}{13}a^{2}x^{3}+\frac{7245}{286}ax^{4}+ \frac{315}{26}x^{5}\right)\\
& & -A_{3}x^{3}\left(\frac{70}{3}a^{5}+\frac{3710}{33}a^{4}x+\frac{2310}{13}a^{3}x^{2}+\frac{17206}{143}a^{2}x^{3}+\frac{392}{13}ax^{4} \right),\\
\Omega(x) &=& A_{1}x\left(84a^{5}+888a^{4}x+3170a^{3}x^{2}+\frac{54490}{11}a^{2}x^{3}+3570ax^{4} +\frac{10710}{11}x^{5} \right)\\
& & +A_{2}x^{2}\left(\frac{165}{2}a^{5}+\frac{1705}{2}a^{4}x+\frac{33505}{11}a^{3}x^{2}+\frac{62474}{13}a^{2}x^{3}+\frac{998235}{286}ax^{4}+\frac{24885}{26}x^{5}\right)\\
& & +A_{3}x^{3}\left(\frac{245}{3}a^{5}+\frac{27475}{33}a^{4}x+\frac{38640}{13}a^{3}x^{2}+\frac{673484}{143}a^{2}x^{3}+\frac{44653}{13}ax^{4}+945x^{5}\right),\\
\end{eqnarray*}
\begin{eqnarray*}
\Lambda(x)&=&A_{1}x\left(252a^{5}+2124a^{4}x+6732a^{3}x^{2}+\frac{100380}{11}a^{2}x^{3}+\frac{63000}{11}ax^{4} +\frac{15120}{11}x^{5} \right)\\
& &+A_{2}x^{2}\left(225a^{5}+1845a^{4}x+\frac{63210}{11}a^{3}x^{2}+\frac{1133370}{143}a^{2}x^{3}+\frac{55755}{11}ax^{4}+\frac{16065}{13}x^{5}\right)\\
& &+A_{3}x^{3}\left(210a^{5}+\frac{18550}{11}a^{4}x+\frac{738360}{143}a^{3}x^{2}+\frac{78624}{11}a^{2}x^{3}+\frac{59934}{13}ax^{4}+1134x^{5} \right).
\end{eqnarray*}
With this exact solution the line element (\ref{line-element}) becomes
\begin{equation*}
\begin{aligned}
ds^{2}=&-A^{2}\left(a+x\right)^{4}dt^{2}+r^{2}(d\theta^{2}+\sin^{2}\theta d\phi^{2})\\
& +\dfrac{{315( a+x)^{2}(a+3x)}dr^{2}}{9\left( 35a^{3}+35a^{2}x+21ax^{2} + 5x^{3} \right)
-\dfrac{2Bx}{C}\left( 105a^{3}+189a^{2}x+135ax^{2} + 35x^{3} \right)+\frac{315 L(x)}{C}}.
\label{nonsingular-line element1}
\end{aligned}
\end{equation*}
The mass function (\ref{mass2}) becomes
\begin{equation*}
\begin{split}
\begin{aligned}
M(x)=& \left(\left(\frac{1268}{96525}a-\frac{1}{30}x^{2}-\frac{14}{585}ax\right) A_{3}-\left(\frac{4}{91}x+\frac{74}{2145}a\right) A_{2}
-\frac{7}{110}A_{1}\right)\frac{x^{\frac{5}{2}}}{C^\frac{3}{2}} \\
&-\left(\frac{1}{9}B+\frac{2}{33}aA_{1}-\frac{10}{429}a^{2}A_{2}+ \frac{14}{1287}a^{3}A_{3}\right)\left(\frac{x}{C}\right)^{\frac{3}{2}}\\
&-\sqrt{\frac{a}{C^{3}}}\left( \frac{62}{105}aB+\frac{93}{35}C+\frac{31}{385}a^{2}A_{1}-\frac{31}{1001}a^{3}A_{2}
+\frac{31}{2145}a^{4}A_{3}\right)\arctan \sqrt{\frac{x}{a}}\\ 
&+\frac{\sqrt{3a}}{3C^\frac{3}{2}}\left( \frac{188}{315}aB+\frac{129}{35}C+\frac{59}{1155}a^{2}A_{1}-\frac{100}{9009}a^{3}A_{2}
+\frac{157}{57915}a^{4}A_{3}\right)\arctan\sqrt{\frac{3x}{a}}\\
&+\left(\frac{76}{189}aB+\frac{8}{9}C+\frac{52}{693}a^{2}A_{1}-\frac{934}{27027}a^{3}A_{2}
+\frac{3088}{173745}a^{4}A_{3}\right)\sqrt{\frac{x}{C^{3}}}\\
&-\left(\frac{6}{35}a^{2}B+\frac{27}{35}aC+\frac{9}{385}a^{3}A_{1}-\frac{9}{1001}a^{4}A_{2}
+\frac{3}{715}a^{5}A_{3}\right)\frac{\sqrt{x}}{(a+x)C^\frac{3}{2}}\\
&-\left(\frac{4}{105}a^{3}B+\frac{6}{35}a^{2}C+\frac{2}{385}a^{4}A_{1}-\frac{2}{1001}a^{5}A_{2}+
\frac{2}{2145}a^{6}A_{3}\right)\frac{\sqrt{x}}{(a+x)^{2}C^{\frac{3}{2}}}\\
&-\left(\frac{188}{945}a^{2}B+\frac{43}{35}aC+\frac{59}{3465}a^{3}A_{2}-\frac{100}{27027}a^{4}A_{2}+
\frac{157}{173745}a^{5}A_{3}\right)\frac{\sqrt{x}}{(a+3x)C^{\frac{3}{2}}}.
\end{aligned}
\end{split}
\end{equation*}


\begin{thebibliography}{}
\bibitem[\protect\citeauthoryear {Astashenok et al.}{2013}]{salvatore2013} Astashenok, A.V., Capozziello, S., Odintsov, S.D.: JCAP {\bf 1312}, 040 (2013)
\bibitem[\protect\citeauthoryear {Bijalwan}{2011}]{Naveen} Bijalwan, N.: Astrophys. Space Sci. \textbf{336}, 413 (2011)
\bibitem[\protect\citeauthoryear {Bombaci}{2000}]{Bombaci} Bombaci, I.: arXiv:astro-ph/0002524v1͒͒ \textbf{} (2000)
\bibitem[\protect\citeauthoryear {Bowers and Liang}{1974}]{Richard} Bowers, R.L., Liang, E.P.T.: Astrophys. J. \textbf{188}, 657 (1974)
\bibitem[\protect\citeauthoryear {Chaisi and Maharaj}{2005}]{Chaisi2} Chaisi, M., Maharaj, S.D.: Gen. Relativ. Gravit. \textbf{37}, 1177 (2005)
\bibitem[\protect\citeauthoryear {Chaisi and Maharaj}{2006a}]{Chaisi} Chaisi, M., Maharaj, S.D.: Pramana - J. Phys. \textbf{66}, 313 (2006a)
\bibitem[\protect\citeauthoryear {Chaisi and Maharaj}{2006b}]{Chaisi3} Chaisi, M., Maharaj, S.D.: Pramana - J. Phys. \textbf{66}, 609 (2006b)
\bibitem[\protect\citeauthoryear {Chattopadhyay et al.}{2012}]{Kumar} Chattopadhyay, P.K., Deb, R., Paul, B.C.: Int. J. Mod. Phys. D \textbf{21}, 1250071 (2012)
\bibitem[\protect\citeauthoryear {Dev and Gleiser}{2002}]{Dev2} Dev, K., Gleiser, M.: Gen. Relativ. Gravit. \textbf{34}, 1793 (2002)
\bibitem[\protect\citeauthoryear {Dev and Gleiser}{2003}]{Dev3} Dev, K., Gleiser, M.: Gen. Relativ. Gravit. \textbf{35}, 1435 (2003)
\bibitem[\protect\citeauthoryear {Dey et al.} {1998}]{Dey} Dey, M., Bombaci I., Dey, J., Ray S., Samanta, B.C.: Phys. Lett. B \textbf{438}, 123 (1998)
\bibitem[\protect\citeauthoryear {Dong et al.} {2013}]{dong13} Dong, H., Kuo, T.T.S., Lee, H.K., Machleidt, R., Rho, M.: Phys Rev C {\bf 87}, 054332 (2013)
\bibitem[\protect\citeauthoryear {Durgapal and Banerjee} {1983}]{Durgapal} Durgapal, M.C., Bannerji, R.: Phys. Rev. D \textbf{27}, 328 (1983)
\bibitem[\protect\citeauthoryear {Esculpi and Aloma}{2010}]{Esculpi} Esculpi, M., Aloma, E.: Eur. Phys. J. C \textbf{67}, 521 (2010)
\bibitem[\protect\citeauthoryear {Feroze and Siddiqui} {2011}]{Feroze} Feroze, T., Siddiqui, A. A.: Gen. Relativ. Gravit. \textbf{43}, 1025 (2011)
\bibitem[\protect\citeauthoryear {Freire et al.} {2011}]{Freire} Freire, P.C.C, Bassa, C.G., Wex, N., Stairs, I.H. \textit{et al}: Mon. Not. R. Astron. Soc. \textbf{412}, 2763 (2011)
\bibitem[\protect\citeauthoryear {Gangopadhyay et al.} {2013}]{Taparati} Gangopadhyay, T., Ray, S., Li, X.D., Dey, J., Dey, M.: Mon. Not. R. Astron. Soc. \textbf{431}, 3216 (2013)
\bibitem[\protect\citeauthoryear {Ganguly et al.} {2014}]{ganguly2014} Ganguly, A., Gannouji, R., Goswami, R., Ray, S.: Phys. Rev D {\bf 89}, 064019 (2014)
\bibitem[\protect\citeauthoryear {Gleiser and Dev} {2004}]{Dev} Gleiser, M., Dev, K.: Int. J. Mod. Phys. D \textbf{13}, 1389 (2004)
\bibitem[\protect\citeauthoryear {Gondek-Rosinaka et al.}{2000}]{dorota2000} Gondek-Rosinska, D., Bulik, T., Gourgoulhon, E., Ray, S., Dey, J., Dey, M.: Astron. Astrophys. {\bf 363}, 1005 (2000)
\bibitem[\protect\citeauthoryear {Gupta and Maurya} {2011a}]{Gupta} Gupta, Y.K., Maurya, S.K.: Astrophys. Space Sci. \textbf{332}, 155 (2011a)
\bibitem[\protect\citeauthoryear {Gupta and Maurya} {2011b}]{Gupta1} Gupta, Y.K., Maurya, S.K.: Astrophys. Space Sci. \textbf{331}, 135 (2011b)
\bibitem[\protect\citeauthoryear {Harko and Mak} {2002}]{Harko} Harko, T., Mak, M.K.: Annalen Phys. \textbf{11}, 3 (2002)
\bibitem[\protect\citeauthoryear {Heinzle et al.} {2003}]{Heinzle} Heinzle, J.M., R\"{o}hr, N., Uggla, C.: Class. Quantum Grav. \textbf{20}, 4567 (2003)
\bibitem[\protect\citeauthoryear {Ivanov} {2010}]{Ivanov2} Ivanov, B.V.: Int. J. Theor. Phys. \textbf{49}, 1236 (2010)
\bibitem[\protect\citeauthoryear {Ivanov} {2002}]{Ivanov} Ivanov, B.V.: Phys. Rev. D \textbf{65}, 104001 (2002)
\bibitem[\protect\citeauthoryear {Kalam et al.} {2012}]{Mehedi} Kalam, M., Rahaman, F., Ray, S., Hossein, Sk.M., Karar, I., Naska, J.: Eur. Phys. J. C \textbf{72}, 2248 (2012)
\bibitem[\protect\citeauthoryear {Kalam et al.} {2013}]{Mehedi1} Kalam, M., Usmani, A.A., Rahamani F., Hossein, S.M., Karar, I., Sharma, R.: Int. J. Theor. Phys. \textbf{52}, 3319 (2013)
\bibitem[\protect\citeauthoryear {Karmakar et al.} {2007}]{Karmakar} Karmakar, S., Mukherjee, S., Sharma, R., Maharaj, S.D.: Pramana - J. Phys. \textbf{68}, 881 (2007)
\bibitem[\protect\citeauthoryear {Kinasiewicz and Mach} {2007}]{Mach} Kinasiewicz, B., Mach, P.: Acta Physica Polonica B \textbf{38}, (2007)
\bibitem[\protect\citeauthoryear {Komathiraj and Maharaj} {2007a}]{Komathiraj2} Komathiraj, K., Maharaj, S.D.: Gen. Relativ. Gravit. \textbf{39}, 2079 (2007a)
\bibitem[\protect\citeauthoryear {Komathiraj and Maharaj} {2007b}]{Komathiraj3} Komathiraj, K., Maharaj, S.D.: J. Math. Phys. \textbf{48}, 042501 (2007b)
\bibitem[\protect\citeauthoryear {Komathiraj and Maharaj} {2007c}]{Komathiraj} Komathiraj, K., Maharaj, S.D.: Int. Mod. Phys. D \textbf{16}, 1803 (2007c)
\bibitem[\protect\citeauthoryear {Lai and Xu} {2009}]{Lai} Lai, X.Y., Xu, R.X.: Astroparticle Phys. \textbf{31}, 128 (2009)
\bibitem[\protect\citeauthoryear {Mafa Takisa and Maharaj} {2013a}]{Mafa} Mafa Takisa, P., Maharaj, S.D.: Astrophys. Space Sci. \textbf{343}, 569 (2013a)
\bibitem[\protect\citeauthoryear {Mafa Takisa and Maharaj} {2013b}]{Mafa2} Mafa Takisa, P., Maharaj, S.D.: Gen. Relativ. Gravit. \textbf{45}, 1951 (2013b)
\bibitem[\protect\citeauthoryear {Maharaj and Chaisi} {2006a}]{Mah-Chaisi} Maharaj, S.D., Chaisi, M.: Gen. Relativ. Gravit. \textbf{38}, 1723, (2006a)
\bibitem[\protect\citeauthoryear {Maharaj and Chaisi} {2006b}]{Maharaj4} Maharaj, S.D, Chaisi, M.: Math. Meth. Appl. Sc. \textbf{29}, 67 (2006b)
\bibitem[\protect\citeauthoryear {Maharaj and Komathiraj} {2007}]{Maharaj3} Maharaj, S.D., Komathiraj, K.: Class. Quantum Grav. \textbf{24}, 4513 (2007)
\bibitem[\protect\citeauthoryear {Maharaj and Mafa Takisa} {2012}]{Maharaj} Maharaj, S.D., Mafa Takisa, P.: Gen. Relativ. Gravit. \textbf{44}, 1419 (2012)
\bibitem[\protect\citeauthoryear {Maharaj and Thirukkanesh} {2009a}]{Maharaj5} Maharaj, S.D., Thirukkanesh, S.: Nonlinear Analysis: RWA \textbf{10}, 3396 (2009)
\bibitem[\protect\citeauthoryear {Maharaj and Thirukkanesh} {2009b}]{Maharaj2} Maharaj, S.D., Thirukkanesh, S.: Pramana - J. Phys. \textbf{72}, 481 (2009)
\bibitem[\protect\citeauthoryear {Maharaj et al.} {2014}]{Jefta1} Maharaj, S.D., Sunzu, J.M., Ray, S.: Eur. Phys. J. Plus \textbf{129}, 3 (2014)
\bibitem[\protect\citeauthoryear {Mak and Harko} {2002}]{Mak5} Mak, M.K., Harko T.: Chin. J. Astron. Astrophys. \textbf{2}, 248 (2002)
\bibitem[\protect\citeauthoryear {Mak and Harko}{2003}]{Mak3} Mak, M.K., Harko, T.: Proc. Roy. Soc. Lond. \textbf{A 459}, 393 (2003)
\bibitem[\protect\citeauthoryear {Mak and Harko} {2004}]{Mak4} Mak, M.K., Harko, T.: Int. J. Mod. Phys. D \textbf{13}, 149 (2004)
\bibitem[\protect\citeauthoryear {Mak and Harko} {2005}]{Mak} Mak, M.K., Harko, T.: Pramana - J. Phys. \textbf{65}, 185 (2005)
\bibitem[\protect\citeauthoryear {Maurya and Gupta} {2012}]{Maurya} Maurya, S.K., Gupta, Y.K.: Phys. Scr. \textbf{86}, 025009 (2012)
\bibitem[\protect\citeauthoryear {Misner and Zapolsky}{1964}]{Misner} Misner, C.W., Zapolsky, H.S.: Phys. Rev. Lett. \textbf{12}, 635 (1964)
\bibitem[\protect\citeauthoryear {Murad and Pant}{2014}]{murad2014} Murad, M.H., Pant, N.:  Astrophys. Space Sci. {\bf 350}, 349 (2014)
\bibitem[\protect\citeauthoryear {Negreiros et al.} {2009}]{Rodrigo} Negreiros, R.P., Weber, F., Malheiro, M., Usov, V.: Phys. Rev. D \textbf{80}, 083006 (2009)
\bibitem[\protect\citeauthoryear {Nilsson and Uggla} {2001}]{Nilsson} Nilsson, U.S., Uggla, C.: Annals Phys. \textbf{286}, 292 (2001)
\bibitem[\protect\citeauthoryear {Rahaman et al.} {2012}]{Farook} Rahaman, F., Sharma, R., Ray, S., Maulick, R., Karar, I.: Eur. Phys. J. C  \textbf{72}, 2071 (2012)
\bibitem[\protect\citeauthoryear {Sharma and Maharaj} {2007}]{Sharma} Sharma, R., Maharaj, S.D.: Mon. Not. R. Astron. Soc. \textbf{375}, 1265 (2007)
\bibitem[\protect\citeauthoryear {Sharma et al.} {2006}]{Sharma2} Sharma, R., Karmakar, S., Mukherjee, S.: Int. J. Mod. Phys. D \textbf{15}, 405 (2006)
\bibitem[\protect\citeauthoryear {Sharma et al.} {2001}]{Sharma3} Sharma, S. Mukherjee, S., Maharaj, S.D.: Gen. Relativ. Gravit. \textbf{33}, 999 (2001)
\bibitem[\protect\citeauthoryear {Shibata} {2004}]{Shibata} Shibata, M.: Astrophys. J. \textbf{605}, 350 (2004)
\bibitem[\protect\citeauthoryear {Sotani and Harada} {2003}]{Sotani1} Sotani, H., Harada, T.: Phys. Rev. D \textbf{68}, 024019 (2003)
\bibitem[\protect\citeauthoryear {Sotani et al.} {2004}]{Sotani2} Sotani, H., Kohri, K., Harada, T.: Phys. Rev. D \textbf{69}, 084008 (2004)
\bibitem[\protect\citeauthoryear {Thirukkanesh and Maharaj} {2006}]{Therukanesh4} Thirukkanesh, S., Maharaj, S.D,: Class. Quantum Grav. \textbf{23}, 2697 (2006)
\bibitem[\protect\citeauthoryear {Thirukkanesh and Maharaj} {2008}]{Thirukkanesh} Thirukkanesh, S., Maharaj, S.D.: Class. Quantum Grav. \textbf{25}, 235001 (2008)
\bibitem[\protect\citeauthoryear {Thirukkanesh and Maharaj}{2009}]{Therukanesh2} Thirukkanesh, S., Maharaj, S.D.: Math. Meth. Appl. Sci. \textbf{32}, 684 (2009)
\bibitem[\protect\citeauthoryear {Thirukkanesh and Ragel} {2012}]{Thirukkanesh3} Thirukkanesh, S., Ragel, F.S.: Pramana - J. Phys. \textbf{78}, 687 (2012)
\bibitem[\protect\citeauthoryear {Tooper} {1964}]{Tooper} Tooper, R.F.: Astrophys. J. \textbf{140}, 434 (1964)
\bibitem[\protect\citeauthoryear {Zdunik} {2000}]{zdunik2000} Zdunik, J.L.: Astron.Astrophys. {\bf 359}, 311 (2000)

\end{thebibliography}
\end{document}